\documentclass[aps,pra,12pt,onecolumn,groupedaddress,nofootinbib]{revtex4}
\usepackage{dsfont}
\usepackage{bbm}
\usepackage[mathscr]{euscript}
\usepackage{amsmath}
\usepackage{amssymb}
\usepackage{amsthm}
\usepackage{amsfonts}
\usepackage{graphicx}
\usepackage{bm}
\usepackage{color}
\usepackage[colorlinks=true,urlcolor=blue,linkcolor=cyan,citecolor=red]{hyperref}
\begin{document}

\title{The Long Neglected Critically Leveraged Portfolio}
\author{M. Hossein Partovi}
\affiliation{Department of Physics and Astronomy, California State
University, Sacramento, California 95819-6041}

\date{\today}

\begin{abstract}

We show that the efficient frontier for a portfolio in which short positions precisely offset the long ones is composed of a pair of straight lines through the origin of the risk-return plane.  This unique but important case has been overlooked because the original formulation of the mean-variance model by Markowitz as well as all its subsequent elaborations have implicitly excluded it by using portfolio weights rather than actual amounts allocated to individual assets.   We also elucidate the properties of portfolios where short positions dominate the long ones, a case which has similarly been obscured by the adoption of portfolio weights.

\end{abstract}
\maketitle

\section*{Introduction}
The mean-variance model of portfolio selection pioneered by Markowitz (1952, 1959) sixty years ago, as well as all its subsequent modifications and extensions, have been formulated in terms of a set of fractions, or portfolio weights, characterizing its composition.  While this procedure is appropriate as well as expedient for all cases where the net capital outlay, which we will refer to as the portfolio ``budget,'' is different from zero, such as in the original formulation of Markowitz where short sales and borrowing were excluded, it does preclude an important case that can occur in the absence of the positivity constraint on portfolio weights.  This is the case of a highly leveraged portfolio where stock purchases are precisely offset by short sales, with the latter represented by negative weights according to present day practice.  While in practice this is an unlikely outcome for most investors because short sales actually require collateral, as Markowitz (1983) was quick to point out, it is nevertheless an important case which highlights the consequences of aggressive leveraging and has unique properties including a distinct frontier.   In fact a simple calculation shows that the zero-budget efficient frontier with short sales allowed is a direct proportionality between the expected portfolio return and standard deviation, i.e., a pair of straight lines meeting at the origin of the risk-return plane.   In other words, the familiar hyperbolic frontier of a finite-budget portfolio degenerates into a pair of straight lines as the budget vanishes.

This communication could have appeared not long after the publication of Markowitz's foundational paper on portfolio selection in March 1952.  Unconstrained weights appeared a mere four months later in Roy (1952), and subsequently resurfaced in Sharpe (1970) and Merton (1972), and in countless other papers since.  They are of course commonplace in present day discussions.  Yet, for reasons already stated, certain aspects of highly leveraged portfolios, including the zero-budget case, have remained unexplored.   These features are simple and basic consequences of the underlying model, and the required analysis is a slight modification of a well known derivation that dates back to Merton (1972).  They are nevertheless important results, not only because of their intrinsic interest, but also due to the historical role of the mean-variance and capital asset pricing models as the bedrock of modern finance theory.

\section{Optimal Portfolio with Short Sales Revisited}

To demonstrate the exceptional nature of the zero-budget optimal portfolio as well as the consequences of aggressive leveraging, we will start with a calculation of the efficient frontier in terms of the actual amounts invested rather than the weights that represent them.   The corresponding optimization problem is then one of determining how a given portfolio budget $\mathcal{B}$ is to be allocated within a set of risky assets, short positions included, in order to achieve a given expected portfolio return $\mathcal{R}$ with the least possible variance.   In symbols,
\begin{equation}
{\mathbf{y}}_{eff}=\arg [\min \mathbf{{y}^{\dag}}\bm{\sigma} \mathbf{y}]\,\, \textrm{s.t.}\,\,
\mathbf{{y}^{\dag}}\mathbf{1}= \mathcal{B}\,\, \textrm{and} \,\, \mathbf{{y}^{\dag}}\mathbf{r}= \mathcal{R}, \label{1}
\end{equation}
where $- \infty < {y}_{i} < +\infty$ and ${r}_{i} > 0 $ represent, respectively, the amount allocated to, and the expected rate of return of, the $i$th asset.  Moreover, $\bm{\sigma}$ is the covariance matrix of the asset set (assumed positive and nonsingular since all assets are presumed to be risky), $\mathbf{1}$ is a vector (or column matrix) all of whose entries are unity, and ``$\dag$'' signifies matrix transposition.  The portfolio budget $\mathcal{B}$ is the net investment outlay and will be unrestricted in magnitude and sign.  We will also treat the expected portfolio return parameter $\mathcal{R}$ as unrestricted, albeit with the understanding that in practice investors will choose portfolios with positive expected returns.  As indicated, the solution to the problem posed in Eq.~(\ref{1}) is ${\mathbf{y}}_{eff}$, the desired optimal allocation vector, with the corresponding value of the portfolio variance $\mathcal{V}$ equal to ${\mathbf{y}}^{\dag}_{eff}\bm{\sigma} {\mathbf{y}}_{eff}$.  Finally, the expected standard deviation for the portfolio, commonly interpreted as portfolio risk, is given by $\Sigma=\sqrt{\mathcal{V}}$.

It should be noted here that $\mathcal{V}$ is positive definite by virtue of the structure of $\bm{\sigma}$ and regardless of the sign of the portfolio budget $\mathcal{B}$.  In other respects, the sign and magnitude of the portfolio budget play an important role, and we will find it convenient to distinguish the three cases where $\mathcal{B}$ is positive, zero, or negative, respectively as ``subcritically," ``critically," or ``supercritically" leveraged.  Thus short positions dominate in a supercritically leveraged portfolio, while long positions are exactly offset by short ones in a critical portfolio.

As long as $\mathcal{B} \neq 0$, the optimization problem in Eq.~(\ref{1}) has an obvious invariance property whereby ${\mathbf{y}}_{eff}$, $\mathcal{R}$, and ${\mathcal{V}}^{\frac{1}{2}}$ scale with $\mathcal{B}$.  Thus it is natural to adopt the portfolio budget as the num\'{e}raire for subcritical portfolios and reformulate the problem in terms of the weights $\mathbf{x}=\mathbf{y}/\mathcal{B}$ and the corresponding rescaled values of expected portfolio return $\mathcal{R}/\mathcal{B}$ and variance $\mathcal{V}/{\mathcal{B}}^{2}$.   The customary practice is to skip these preliminaries and start with the rescaled version \textit{ab initio}.  Needless to say, the rescaling transformation is ill-defined for $\mathcal{B}=0$, while the underlying problem is of course well posed for all values of $\mathcal{B}$.  This is the long-overlooked case of the efficient frontier alluded to earlier.   Moreover, in case of supercritical portfolios where $\mathcal{B} < 0$ , the use of weights amounts to using a negative quantity as num\'{e}raire.  Aside from being unpalatable, a negative scale is nonintuitive and serves to obscure important features of highly leveraged portfolios.  In addition, one must carefully consider the sign reversals that result from the rescaling, as will be seen later.  These basic consequences of rescaling are often ignored when discussing the efficient frontier, in part because such discussions tacitly assume a positive budget and focus on the properties of subcritical portfolios.

In order to highlight the foregoing points, we will analyze the problem defined in Eq.~(\ref{1}) in terms of the total budget, expected return, and variance, i.e., without rescaling.  Following standard procedure, we introduce the three coefficients
\begin{equation}
{A}_{11}=\mathbf{{1}^{\dag}}{\bm{\sigma}^{-1}} \mathbf{1};\,\,\,{A}_{1r}=\mathbf{{1}^{\dag}}\bm{{\sigma}^{-1}} \mathbf{r};\,\,\, {A}_{rr}=\mathbf{{r}^{\dag}}{\bm{\sigma}^{-1}} \mathbf{r}; \,\,\,D={A}_{11}{A}_{rr}-{{A}_{1r}}^{2}, \label{2}
\end{equation}
in terms of which we obtain the following solution to Eq.~(\ref{1}):
\begin{equation}
{\mathbf{y}}_{eff}=\frac{{A}_{rr}\mathcal{B}-{A}_{1r}\mathcal{R}}{D}\bm{{\sigma}^{-1}}\mathbf{1}+
\frac{{A}_{11}\mathcal{R}-{A}_{1r}\mathcal{B}}{D}{\bm{\sigma}^{-1}}\mathbf{r}. \label{3}
\end{equation}
We pause to note here that $\bm{{\sigma}^{-1}}$ is positive and nonsingular, so that ${A}_{11}$ and ${A}_{rr}$ are positive by virtue of their quadratic structure and $D \geq 0$ by the Cauchy-Schwarz inequality.

The mean-variance characteristics of the optimal portfolios are then given by
\begin{equation}
\mathcal{V}={\mathcal{V}}_{min}+\frac{{A}_{11}}{D}{(\mathcal{R}-{\mathcal{R}}_{min})}^{2}, \label{4}
\end{equation}
where ${\mathcal{V}}_{min}$, which equals ${\mathcal{B}}^{2}/{A}_{11}$, represents the minimum variance portfolio, and ${\mathcal{R}}_{min}$, which equals ${{A}_{1r}}\mathcal{B}/{{A}_{11}}$, is the corresponding value of the expected portfolio return.  Note that ${\mathcal{R}}_{min}$ vanishes for critical portfolios, and is negative for supercritical ones.

The efficient frontier is found from Eq.~(\ref{4}) by solving for $\mathcal{R}$;
\begin{equation}
\mathcal{R}=\frac{{A}_{1r}}{{A}_{11}}\mathcal{B}\pm
{\big[({A}_{rr}-\frac{{{A}_{1r}}^{2}}{{A}_{11}})({\mathbf{\Sigma}}^{2}-\frac{1}{{A}_{11}}{\mathcal{B}}^{2})\big]}^{\frac{1}{2}}, \label{5}
\end{equation}
where $\Sigma=\sqrt{\mathcal{V}}$ is the standard deviation of the optimal portfolio. As stated earlier, investors will choose the portfolio corresponding to the upper sign in Eq.~(\ref{5}) regardless of the sign or magnitude of $\mathcal{B}$.   We will nevertheless continue to include both branches in the efficient frontier.
\subsection{Critically Leveraged Portfolio}
For a critically leveraged portfolio, $\mathcal{B}=0$ and Eqs.~(\ref{3}) and (\ref{5}) reduce to
\begin{equation}
{\mathbf{y}}_{eff}=\mathcal{R}\frac{{\bm{\sigma}^{-1}}({A}_{11}\mathbf{r}-{A}_{1r}\bm{\mathbf{1}})}{D},   \label{6}
\end{equation}
and
\begin{equation}
\mathcal{R}=\pm {({A}_{rr}-\frac{{{A}_{1r}}^{2}}{{A}_{11}})}^{\frac{1}{2}}{\mathbf{\Sigma}}, \label{7}
\end{equation}
respectively.

The first of these, Eq.~(\ref{6}), defines a portfolio with a unique composition whose overall scale and sign are determined by the expected portfolio return $\mathcal{R}$, with the portfolio vanishing altogether as $\mathcal{R}\rightarrow 0$.  The \textit{relative} weights for the portfolio are conveniently defined by using ${\mathcal{R}}$ as the num\'{e}raire;
\begin{equation}
\frac{{\mathbf{y}}_{eff}}{\mathcal{R}}=\frac{{\bm{\sigma}^{-1}}({A}_{11}\mathbf{r}-{A}_{1r}\bm{\mathbf{1}})}{D}.   \label{8}
\end{equation}
This equation underscores the universal nature of the critically leveraged optimal portfolio.  Note that these relative weights add up to zero, not unity as required of proper weights.  Note also that reversing the sign of $\mathcal{R}$ simply exchanges the long and short positions in the portfolio.  Clearly, the critically leveraged portfolio defines a universal allocation model obeying a one-mutual-fund theorem.

The efficient frontier for the critically leveraged portfolio, Eq.~(\ref{7}), is a simple proportionality between risk and return.  Each branch of the frontier is a straight line starting at the origin.  Note that the origin itself corresponds to the null portfolio with ${\mathbf{y}}_{eff}=\mathbf{0}$.  The familiar hyperbolic frontier has thus degenerated into a pair of straight lines as can be seen in Fig.~\ref{fig1}.  This and other features of efficient frontier are best scrutinized in relation to other portfolio types which we consider next.
\subsection{Efficient Frontier in terms of Total Risk-Return Variables}
\begin{figure}
\includegraphics[]{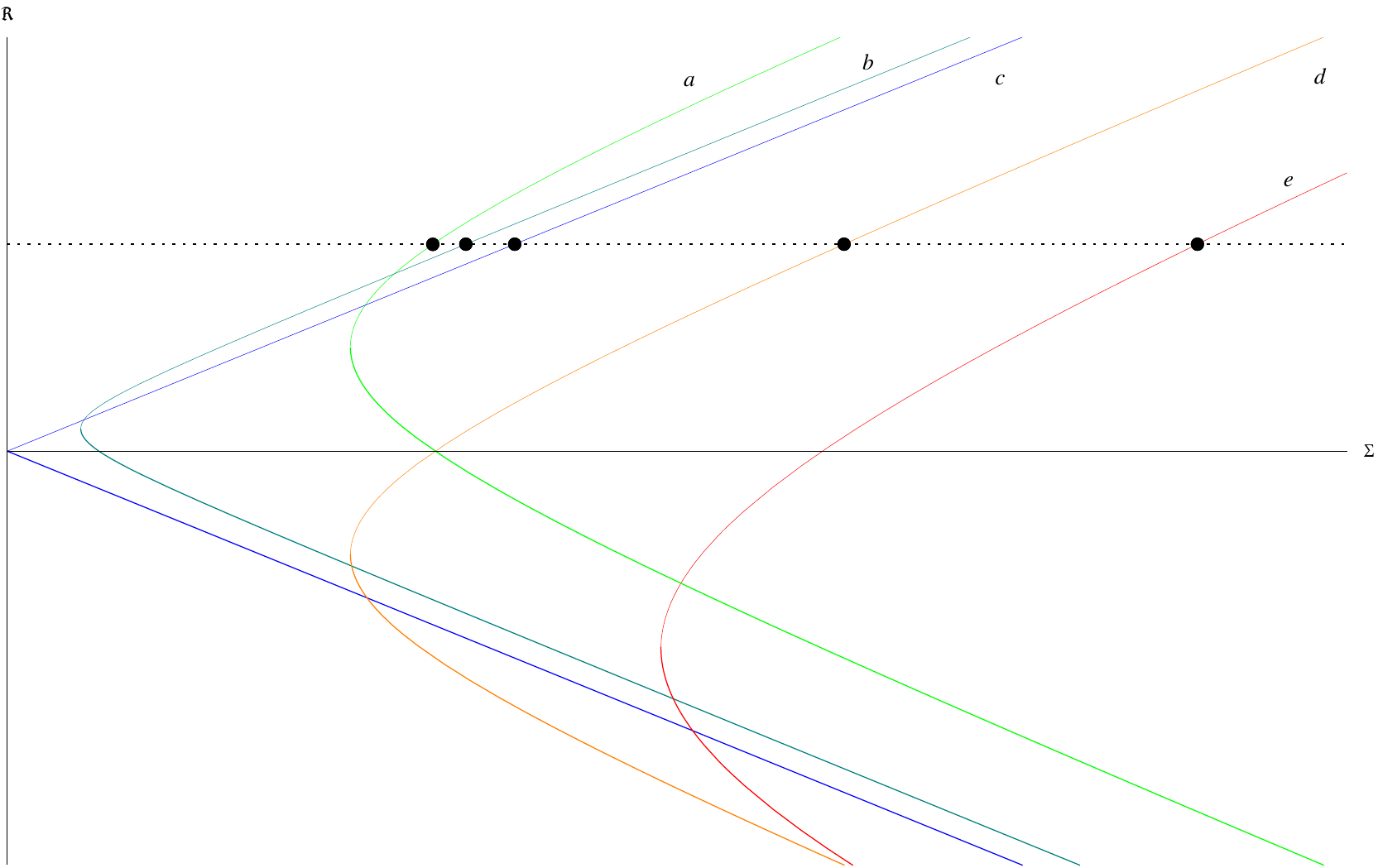}
\caption{Efficient frontier, Eq.~(\ref{5}), for five different values of the portfolio budget arranged in decreasing order: (a) positive, (b) positive (c) zero, (d) negative, and (e) negative.  Portfolios in (a) and (d) differ in the sign of the budget only.   The intersection of the constant return (dotted) line with the five frontiers illustrates the increase in risk as the portfolio budget is reduced.}
\label{fig1}
\end{figure}

Several instances of the efficient frontier in terms of total portfolio variables as given in Eq.~(\ref{5}) are plotted in Fig.~\ref{fig1}.  The particular features of the three portfolio types defined in \S I are clearly in evidence here.  The five cases plotted are labeled (a) through (e) and arranged in order of decreasing portfolio budget $\mathcal{B}$, with (c) representing the critically leveraged portfolio, $\mathcal{B}=0$.  Cases (a) and (d) have budgets of equal magnitude and opposite sign, and serve to illustrate the effect of reversing the sign of $\mathcal{B}$.  The dotted line representing a constant portfolio return, on the other hand, intersects the frontiers of decreasing $\mathcal{B}$ at increasing values of $\Sigma$, showing that portfolio risk rises as its budget is reduced.
As noted earlier, the critical frontier, plot (c), is represented by straight lines starting at the origin of the risk-return plane.  This distinct behavior corresponds to the singular limit of $\mathcal{B}\rightarrow 0$, which shows that the familiar hyperbolic form of the frontier degenerates into a pair of straight lines at the limit $\mathcal{B}=0$.  Plot (b), which corresponds to a subcritical portfolio with a "small" value of $\mathcal{B}$, suggests a continuous transition.

Note that the minimum risk optimal portfolio (leftmost point of each frontier) has the same sign as the budget, and illustrates the fact that negative budgets require elevated risk levels in order to ensure a positive expected return.   It is important to remark here that in comparing the frontiers in Fig.~\ref{fig1} the opportunity cost of the portfolio budget $\mathcal{B}$, which represents the total capital outlay of the investor, is not represented and must be accounted for separately.  In particular, negative budget portfolios have negative opportunity cost, i.e., the funds that (in the simplified model of short sales employed here) the investor receives in return for short positions can be rented out for profit.  As remarked in the introduction, critical and supercritical portfolios are primarily theoretical constructs that serve to underscore the risk-return behavior of highly leveraged portfolios.  They do not represent realistic investment strategies for typical investors.
\subsection{Efficient Frontier in terms of Portfolio Weights}
To make contact with the customary analysis of the unconstrained mean-variance model, essentially as formulated by Merton (1972), we will adopt the portfolio budget $\mathcal{B}$ as num\'{e}raire and rescale Eq.~(\ref{5}) as follows:
\begin{equation}
{r}=\frac{{A}_{1r}}{{A}_{11}}\pm \frac{\mathcal{B}}{|\mathcal{B}|}
{\big[({A}_{rr}-\frac{{{A}_{1r}}^{2}}{{A}_{11}})({\sigma}^{2}-\frac{1}{{A}_{11}})\big]}^{\frac{1}{2}}\,\,(\mathcal{B}\,\,\mathrm{is\,\, num\acute{e}raire}),  \label{9}
\end{equation}
where $r=\mathcal{R}/\mathcal{B}$ and ${\sigma}=\Sigma/|\mathcal{B}|$ are the rescaled values of expected portfolio return and risk.  The portfolio selection protocol is then given by
\begin{equation}
{\mathbf{x}}_{eff}=\frac{{A}_{rr}-{A}_{1r}r}{D}\bm{{\sigma}^{-1}}\mathbf{1}+
\frac{{A}_{11}r-{A}_{1r}}{D}{\bm{\sigma}^{-1}}\mathbf{r}\,\,\,(\mathcal{B}\,\,\mathrm{is\,\, num\acute{e}raire}), \label{10}
\end{equation}
where ${\mathbf{x}}_{eff}={\mathbf{y}}_{eff}/\mathcal{B}$ with the constraint conditions $\mathbf{{x}^{\dag}}\mathbf{1}=1$ and $\mathbf{{x}^{\dag}}\mathbf{r}= r$.  These results are the familiar ones except for the appearance of ${\mathcal{B}}/{|\mathcal{B}|}$, which is the sign of the portfolio budget, in Eq.~(\ref{9}).  For ${\mathcal{B}}>0$, this factor equals unity and can be disregarded, while for ${\mathcal{B}}=0$ it signals the breakdown of the rescaling transformation.  For  ${\mathcal{B}}<0$, on the other hand, the sign factor flips the two branches of the efficient frontier and in addition reverses the overall sign of $r$, a transformation that maps plot (a) to plot (d) of Fig.~\ref{fig1}.  The latter of course simply reflects the fact that an investor with mostly short positions is looking for price drops and negative returns.

A mathematically preferable choice of the num\'{e}raire for rescaling Eq.~(\ref{5}) is $|\mathcal{B}|$, which leads to
\begin{equation}
{r}=\frac{\mathcal{B}}{|\mathcal{B}|}\frac{{A}_{1r}}{{A}_{11}}\pm
{\big[({A}_{rr}-\frac{{{A}_{1r}}^{2}}{{A}_{11}})({\sigma}^{2}-\frac{1}{{A}_{11}})\big]}^{\frac{1}{2}}\,\,(|\mathcal{B}|\,\,\mathrm{is\,\, num\acute{e}raire}),  \label{11}
\end{equation}
and
\begin{equation}
{\mathbf{x}}_{eff}=\frac{{A}_{rr}{\mathcal{B}}/{|\mathcal{B}|}-{A}_{1r}r}{D}\bm{{\sigma}^{-1}}\mathbf{1}+
\frac{{A}_{11}r-{A}_{1r}{\mathcal{B}}/{|\mathcal{B}|}}{D}{\bm{\sigma}^{-1}}\mathbf{r}\,\,\,(|\mathcal{B}|\,\,\mathrm{is\,\, num\acute{e}raire}), \label{12}
\end{equation}
where now ${\mathbf{x}}_{eff}={\mathbf{y}}_{eff}/|\mathcal{B}|$, $r=\mathcal{R}/|\mathcal{B}|$, and ${\sigma}=\Sigma/|\mathcal{B}|$ are the rescaled quantities.  The constraint conditions are now $\mathbf{{x}^{\dag}}\mathbf{1}={\mathcal{B}}/{|\mathcal{B}|}$ and $\mathbf{{x}^{\dag}}\mathbf{r}= r$.  While perfectly logical and substantively equivalent to the formulation of Eqs.~(\ref{9}) and (\ref{10}), this version is even less desirable because of the appearance of negative ``weights'' and other sign reversals for ${\mathcal{B}}<0$.   Overall, both rescaled versions are problematic when dealing with negative budget portfolios.
\section{Concluding Remarks}
It is clear that the traditional formulations of the efficient frontier in terms of portfolio weights fail to provide a satisfactory characterization of the optimal portfolio in the case of critical and supercritical portfolios.  For these highly leveraged portfolio types, the more general formulation of the previous subsections, specifically \S IB, afford a complete and unambiguous description.

\eject

\section*{References}

\noindent
Markowitz, Harry, 1952, Portfolio selection, {\it Journal of
Finance} 7, 77-91.\\
Markowitz, Harry M., 1959, {\it Portfolio Selection: Efficient
Diversification of Investments} (John Wiley \& Sons, New York, NY.).\\
Markowitz, Harry, 1983, Nonnegative or Not Nonnegative: A Question about CAPMs, {\it Journal of
Finance} 38, 283-295.\\
Merton, Robert, 1972, An Analytic Derivation of the Efficient
Portfolio Frontier, {\it Journal of Finance and Quantitative
Analysis} 7, 1851-1872.\\
Roy, Andrew D., 1952, Safety First and the Holding of Assets, {\it
Econometrica} 34, 431-449.\\
Sharpe, William F., 1970, \textit{Portfolio Theory and Capital Markets} (McGraw-Hill, New York, NY.).\\

\end{document}